\documentclass[12pt,a4paper,english]{article}
\usepackage[T1]{fontenc}
\usepackage[latin9]{inputenc}
\usepackage{float}
\usepackage{amsmath}
\usepackage{amssymb}

\makeatletter

\providecommand{\tabularnewline}{\\}

\makeatother

\usepackage{babel}

\begin{document}
\thispagestyle{empty}

\begin{center}
\textbf{\LARGE Factorization Properties of}
\par\end{center}{\LARGE \par}

{\LARGE \smallskip{}
}{\LARGE \par}

\begin{center}
\textbf{\LARGE Finite Spaces}
\par\end{center}{\LARGE \par}

\bigskip{}

\begin{center}
\textbf{\large B Simkhovich$^{1}$, A Mann$^{1,2}$ and J Zak$^{1}$}
\par\end{center}{\large \par}

{\scriptsize \bigskip{}
}{\scriptsize \par}

\begin{center}
{\scriptsize $^{1}$Department of Physics, Technion - Israel Institute
of Technology, Haifa 32000, Israel}\\
{\scriptsize $^{2}$Department of Physics, National Cheng Kung
University, Tainan 70101, Taiwan}
\par\end{center}{\scriptsize \par}

{\scriptsize \smallskip{}
}{\scriptsize \par}

\begin{center}
{\scriptsize E-mail: boriskas@tx.technion.ac.il, ady@physics.technion.ac.il,
zak@physics.technion.ac.il}
\par\end{center}{\scriptsize \par}

\pagestyle{plain} 
\begin{abstract}
\setcounter{page}{1} 

In 1960 Schwinger {[}{\footnotesize J. Schwinger, }\textsl{\footnotesize Proc.Natl.Acad.Sci.}{\footnotesize{}
}\textbf{\textsl{\footnotesize 46}}{\footnotesize{} }\textsl{\footnotesize (1960)
}{\footnotesize 570-579}{]} proposed the algorithm for factorization
of unitary operators in the finite $M$ dimensional Hilbert space
according to a coprime decomposition of $M$. Using a special permutation
operator $A$ we generalize the Schwinger factorization to every decomposition
of $M$. We obtain the factorized pairs of unitary operators and show
that they obey the same commutation relations as Schwinger's. We apply
the new factorization to two problems. First, we show how to generate
two \textit{kq}-like mutually unbiased bases for any composite dimension.
Then, using a Harper-like Hamiltonian model in the finite dimension
$M=M_{1}M_{2}$, we show how to design a physical system with $M_{1}$
energy levels, each having degeneracy $M_{2}.$
\end{abstract}
$\numberwithin{equation}{section}$

\section{Introduction}

A finite phase space of dimension $M$, where coordinate and momentum
have $M$ possible values, is a frequent component of various physical
and mathematical problems. Fast Fourier Transform (FFT) \cite{Good fft,C-T},
Schwinger factorization of unitary operators \cite{Sch 1}, generation
of \textit{kq} bases and finite dimensional Harper-like Hamiltonians
\cite{MRZ} are the problems related to finite phase space that will
be considered in this paper. A recent review of various quantum systems
with finite Hilbert space can be found in ref. \cite{Vourdas}.

Originally studied by Weyl \cite{Weyl}, the finite dimensional Hilbert
space was systematized by Schwinger in terms of {}``Unitary Operator
Bases'' \cite{Sch 1}. Schwinger considered a $M$-dimensional physical
system. Such a Hilbert space can be achieved by application of the
following boundary conditions on the wave function $\psi(x)$ and
its Fourier transform $\Psi(p)$ (ref. \cite{Zak boundary}):

\begin{equation}
\psi(x)=\psi(x+Mc),\,\,\Psi(p)=\Psi(p+\frac{2\pi\hbar}{c}),\end{equation}
where $c$ is a length unit. In what follows, we will assume $c=1$.
As a consequence of the above boundary conditions $x$ and $p$ have
a finite discrete spectrum of eigenvalues:

\begin{equation}
x=0,1,...,M-1;\,\, p=\frac{2\pi\hbar}{M}\cdot\{0,1,...,M-1\}.\end{equation}
Using unitary operators $U$ and $V$ (ref. \cite{MRZ} with $c=1$):

\begin{equation}
\left\{ \begin{array}{l}
U=e^{i\hat{x}\frac{2\pi{}}{M}},\\
V=e^{\frac{i}{\hbar}\hat{p}},\end{array}\right.\label{eq:UV}\end{equation}
the complete orthogonal operator basis of $M^{2}$ operators can be
defined as \cite{Sch 1}:

\begin{equation}
U^{k}V^{n};\, k,n=0,1,...,M-1.\label{eq:Sch Bases}\end{equation}
The above operators have the commutation relation:

\begin{equation}
V^{n}U^{k}=U^{k}V^{n}e^{\frac{2\pi i}{M}nk}.\label{eq:U-V comm}\end{equation}
For a coprime decomposition of $M=M_{1}M_{2}$, using the Fermat-Euler
theorem, Schwinger showed how to factorize the unitary operators.
The Fermat-Euler theorem states that if $M_{1}$ and $M_{2}$ are
coprime, then there exist unique $N_{1}$ and $N_{2}$ such that:

\begin{equation}
M_{1}N_{2}=1\,(\text{mod }M_{2}),\,\, M_{2}N_{1}=1\,(\text{mod }M_{1}).\end{equation}
Therefore, the two pairs of unitary operators defined as:

\begin{equation}
\left\{ \begin{array}{l}
U_{1}=U^{M_{2}},\\
V_{1}=V^{M_{2}N_{1}},\end{array}\right.\left\{ \begin{array}{l}
U_{2}=U^{M_{1}},\\
V_{2}=V^{M_{1}N_{2}},\end{array}\right.\label{eq:Sch U V}\end{equation}
behave as independent complementary operators of factorized dimensions
$M_{1}$ and $M_{2}$. The respective commutation relations are:

\begin{equation}
V_{i}^{n_{i}}U_{i}^{k_{i}}=U_{i}^{k_{i}}V_{i}^{n_{i}}e^{\frac{2\pi i}{M_{i}}n_{i}k_{i}},\, i=1,2;\label{eq:Schw U V 1}\end{equation}

\begin{equation}
V_{i}^{n_{i}}U_{j}^{k_{j}}=U_{j}^{k_{j}}V_{i}^{n_{i}},\, i\neq j,\, i,j=1,2.\label{eq:Schw U V 2}\end{equation}

Each unitary operator on the $M$ dimensional space (Eq. \ref{eq:Sch Bases})
can be considered as two operators from the factorized dimensions
$M_{1}$ and $M_{2}$. This is due to the one-to-one {}``Sino-Ruritanian''
correspondences \cite{Good 2FFT}:

\begin{equation}
\begin{split}\begin{array}{c}
n=n_{1}M_{2}N_{1}+n_{2}M_{1}N_{2}\,\,\,(mod\, M),\\
k=k_{1}M_{2}+k_{2}M_{1}\,\,\,(mod\, M).\end{array}\end{split}
\label{eq:Sino-Ru}\end{equation}
Therefore, for every power $n$ of the operator $V$ we can find the
unique representation by the factorized unitary operators $V_{1}$
and $V_{2}$. The appropriate powers $n_{1}$ and $n_{2}$ of the
factorized operators $V_{1}$ and $V_{2}$ are determined by the first
{}``Sino-Ruritanian'' correspondence (Eq. \eqref{eq:Sino-Ru}).
Similarly, the correspondence between $U$ and $(U_{1},U_{2})$ is
determined by the second {}``Sino-Ruritanian'' correspondence (Eq.
\eqref{eq:Sino-Ru}). Another recent factorization construction based
on the Chinese Remainder Theorem (CRT) can be found in ref. \cite{Vourdas}.

After we have obtained factorization of the $M$ dimensional Hilbert
space into its coprime sub-dimensions $M_{1}$ and $M_{2}$, we can
apply it to the \textit{kq} bases generation and the Harper-like Hamiltonian
model. Let us first consider the \textit{kq} bases generation. The
factorized operators from Eq. \eqref{eq:Sch U V} can be used for
generation of the following two pairs of operators (note that $V_{2}^{M_{1}}=V^{M_{1}}$
and $V_{1}^{M_{2}}=V^{M_{2}}$):

\begin{equation}
\begin{split}(a)\left\{ \begin{array}{l}
\tau(\frac{2\pi{}}{a})=e^{i\hat{x}\frac{2\pi{}}{a}}=U^{M_{2}};\\
T(a)=e^{\frac{i}{\hbar}\hat{p}a}=V^{M_{1}};\end{array}\right.(b)\left\{ \begin{array}{l}
\tau(\frac{2\pi{}}{b})=e^{i\hat{x}\frac{2\pi{}}{b}}=U^{M_{1}};\\
T(b)=e^{\frac{i}{\hbar}\hat{p}b}=V^{M_{2}},\end{array}\right.\end{split}
\label{eq:Zak operators}\end{equation}
where the dimension $M=M_{1}M_{2}$ and $a=M_{1}$, $b=M_{2}$ (according
to the notation of ref. \cite{MRZ} with c=1). Hence, by employing
all possible powers, each pair of operators (a) and (b) forms a complete
set of M commuting operators and thus generates an alternative \textit{kq}
basis for treatment of the $M$-dimensional Hilbert space. We have
two such bases:

\begin{gather}
\begin{split}\begin{array}{c}
(a)\text{ }|k,q\rangle=\frac{1}{\sqrt{M_{2}}}\sum_{s=0}^{M_{2}-1}e^{iksa}|q+sa\rangle,\\
\\(a)\left\{ \begin{array}{l}
k=\frac{2\pi}{M}f,\,\text{ }f=0,...,M_{2}-1,\\
q=0,...,M_{1}-1,\end{array}\right.\end{array}\end{split}
\begin{split}\begin{array}{cc}
(b)\text{ }|K,Q\rangle=\frac{1}{\sqrt{M_{1}}}\sum_{t=0}^{M_{1}-1}e^{iKtb}|Q+tb\rangle\\
\\(b)\left\{ \begin{array}{l}
K=\frac{2\pi}{M}f',\,\text{ }f'=0,...,M_{1}-1,\\
Q=0,...,M_{2}-1.\end{array}\right.\end{array}\end{split}
\label{eq:kq}\end{gather}
The unique property of the \textit{kq} bases is that they are eigenfunctions
of both space and momentum displacement operators. These functions
have partial knowledge about both position and momentum, whose precise
simultaneous knowledge is limited by the non-commutation of the corresponding
operators. In the case of dimension $M=M_{1}M_{2}$ factorizable to
coprime numbers $M_{1}$ and $M_{2}$, the two \textit{kq} bases (a)
and (b) are Mutually Unbiased Bases (MUB) \cite{MRZ}. The MUB property
of the bases means that if the physical system is found in one of
the states of one MUB (for example set (a)), then it has equal probabilities
to be in all the states of the other MUB (set (b) in our example).
Mathematically, the mutual unbiasedness of the two \textit{kq} bases
means the following equality: $|\langle k,q|K,Q\rangle|^{2}=\frac{1}{M}$.
For non-coprime $M_{1}$ and $M_{2}$ the MUB property is violated.
For example, if $M_{1}=m_{1}r$ and $M_{2}=m_{2}r$ have a common
multiple $r$, we have:

\begin{equation}
\langle k,q|K,Q\rangle=\frac{1}{\sqrt{M}}\sum_{t,s}e^{-iksa}e^{iKtb}\langle sm_{1}r+q|tm_{2}r+Q\rangle.\end{equation}
The product $\langle sm_{1}r+q|tm_{2}r+Q\rangle$ equals unity for
the solution of the following modular equation:

\begin{equation}
sm_{1}r+q-tm_{2}r-Q=0\,(mod\, M).\end{equation}
Following ref. \cite{Vinogradov} (p. 45 theorem `d'), the above equation
can be taken modulo $r$:

\begin{equation}
q=Q\,(mod\, r).\end{equation}
Therefore, for $q_{0}=0$ and $Q_{0}=1$ (which is always possible
according to the ranges of values Eqs. (\ref{eq:kq})) we have $|\langle k,q_{0}|K,Q_{0}\rangle|^{2}=0\neq\frac{1}{M}$.

To complete the introduction to \textit{kq} MUB we note their quasi-periodic
properties:

\begin{equation}
\begin{split}\begin{array}{cc}
(a)\,|k+\frac{2\pi}{M_{1}},q\rangle=|k,q\rangle,\, & |k,q+M_{1}\rangle=e^{-ika}|k,q\rangle,\\
(b)\,|K+\frac{2\pi}{M_{2}},Q\rangle=|K,Q\rangle,\, & |K,Q+M_{2}\rangle=e^{-iKb}|K,Q\rangle.\end{array}\end{split}
\label{eq:Quasi kq}\end{equation}

Now, let us consider Harper-like Hamiltonians. They are defined as
Hamiltonians of one degree of freedom periodic both in coordinate
and momentum \cite{Harper-like}. For our discussion we are interested
in the use of Harper-like Hamiltonians for the energy spectra design
considered in ref. \cite{MRZ}. The energy spectra design is a direct
consequence of the factorization of the $M=M_{1}M_{2}$ - dimensional
Hilbert space to coprime constituents $M_{1}$ and $M_{2}$. In the
original version (ref. \cite{MRZ}) one considered a Harper-like Hamiltonian
$H[T(b),\tau(\frac{2\pi}{a})]$ which is a function of the two operators
$T(b)$ and $\tau(\frac{2\pi}{a})$ from Eq. \eqref{eq:Zak operators}.
It is important that the Hamiltonian $H[T(b),\tau(\frac{2\pi}{a})]$
is a function of the operators $V_{1}$ and $U_{1}$ (due to $T(b)=V_{1}^{M_{2}}$
and $\tau(\frac{2\pi}{a})=U_{1}$); in such a case only the $M_{1}$
- dimensional subspace is affected by the Hamiltonian. The $M_{2}$
- dimensional subspace is untouched by the Hamiltonian. Hence, considering
$H[T(b),\tau(\frac{2\pi}{a})]$ we expect to obtain $M_{1}$ energy
levels (with a spectrum determined by the details of the Hamiltonian)
each of which is degenerate $M_{2}$ times.

The aim of this paper is first to extend the Schwinger factorization
to non-coprime $M_{1}$ and $M_{2}$. Then the other two applications,
the \textit{kq}-like MUB generation and the energy spectra design
by Harper-like Hamiltonian, are extended correspondingly. For that
purpose, in section 2, we define the permutation operator $A$, based
on the previous study by Cooley and Tukey of Fast Fourier Transform
(FFT) \cite{C-T}. Using the operator $A$ we obtain pairs of unitary
operators, which have commutation relations as in Eqs. (\ref{eq:Schw U V 1},
\ref{eq:Schw U V 2}). In section 3 we use the new factorized unitary
pairs to generate two \textit{kq}-like MUB. New quasi-periodicity
properties are obtained in one of the bases. In section 4 we apply
the new factorization to the energy spectra design using Harper-like
Hamiltonians without any restriction on the factors $M_{1}$ and $M_{2}$
of the dimension $M=M_{1}M_{2}$. Section 5 includes a discussion
and summary.

\section{Factorization of unitary operators using the permutation operator
A}

To define the permutation operator $A$ we start by recalling the
Division Algorithm Theorem (DAT) from number theory.

The theorem states (ref. \cite{Vinogradov} page 2 or ref. \cite{Natha}
page 3) that for any integer numbers $D$ and $d$ with $d>0$, there
exists a unique pair of integer numbers $q$ and $r$ satisfying the
following conditions: \begin{equation}
\begin{split}\begin{array}{l}
(a)\text{ }D=d\cdot{q}+r,\\
(b)\text{ }0\leq{r}<d.\end{array}\end{split}
\end{equation}

Consider the special case of positive integer $D$ in the range $[0,1,...,M_{1}M_{2}-1]$
and positive $d=M_{2}$. In this case there is a unique pair of integers
$q$ and $r$ satisfying the following conditions: \begin{equation}
\begin{split}\begin{array}{l}
(a)\text{ }q\in{[0,1,...,M_{1}-1]},\\
(b)\text{ }r\in{[0,1,...,M_{2}-1]},\\
(c)\text{ }D=d\cdot{q}+r.\end{array}\end{split}
\end{equation}
For our discussion, this DAT based special representation of the numbers
modulo $M=M_{1}M_{2}$ is the crucial component.

In 1965 Cooley and Tukey \cite{C-T} introduced an FFT algorithm not
limited to coprime factorization of $M=M_{1}M_{2}$. They used two
complementary DAT based representations for the $x$ and $p$ variable
indices:

\begin{equation}
\begin{split}\left\{ \begin{array}{l}
n=n_{1}M_{2}+n_{2},\\
k=k_{1}+k_{2}M_{1},\end{array}\right.\end{split}
\label{eq:conj ENR}\end{equation}
which enabled them to simplify the Discrete Fourier Transform (DFT)
calculation ($w_{M}=e^{\frac{2\pi{}i}{M}}$):

\begin{equation}
\begin{split}\begin{array}{c}
p_{k}=\frac{1}{\sqrt{M}}\sum_{n=0}^{M-1}x_{n}w_{M}^{nk}=\frac{1}{\sqrt{M}}\sum_{n=0}^{M-1}x_{n}w_{M}^{(n_{1}M_{2}+n_{2})(k_{1}+k_{2}M_{1})}=\\
\\=\frac{1}{\sqrt{M_{2}}}\sum_{n_{2}=0}^{M_{2}-1}w_{M}^{n_{2}(k_{1}+k_{2}M_{1})}\frac{1}{\sqrt{M_{1}}}\sum_{n_{1}=0}^{M_{1}-1}x_{n}w_{M}^{n_{1}k_{1}M_{2}}.\end{array}\end{split}
\label{eq:C-T FFT}\end{equation}
The two summations in the last line of the above formula require $M\cdot\sum_{i=1}^{2}M_{i}$
operations instead of $M^{2}$ operations by direct calculation \cite{C-T}.

Following Cooley and Tukey, we define a permutation operator $A$,
which acts in the finite $M$ - dimensional Hilbert space:

\begin{equation}
A=\sum_{x_{1}=0}^{M_{1}-1}\sum_{x_{2}=0}^{M_{2}-1}|x_{2}+M_{2}x_{1}\rangle\langle x_{1}+M_{1}x_{2}|.\label{eq:A operator}\end{equation}
For the construction of the operator $A$ we used two DAT based representations,
as in Eq. \eqref{eq:conj ENR}, applied to the coordinate states $|x\rangle$.
In the coordinate representation our operator is equal to the stride
permutation matrix widely used in signal processing \cite{Toli-An}.
For a simple illustration, let us consider the example of dimension
$M=6$, where $M_{1}=2$ and $M_{2}=3$. The table of correspondence
between the numbers $x=0,1,2,3,4,5$ and pairs of numbers $(x_{1}=0,1;\, x_{2}=0,1,2)$
according to the rule $x=x_{1}+M_{1}x_{2}$ is:

\begin{table}[H]
\begin{centering}
\begin{tabular}{|c|c|c|c|c|c|c|}
\hline 
$x$ & 0 & 1 & 2 & 3 & 4 & 5\tabularnewline
\hline
\hline 
$x_{1}$ & 0 & 1 & 0 & 1 & 0 & 1\tabularnewline
\hline 
$x_{2}$ & 0 & 0 & 1 & 1 & 2 & 2\tabularnewline
\hline
\end{tabular}
\par\end{centering}

\caption{DAT representation of $x=(x_{1},x_{2})$}

\end{table}
With the rule $x=x_{1}^{'}M_{2}+x_{2}^{'}$ we have another table
:

\begin{table}[H]
\begin{centering}
\begin{tabular}{|c|c|c|c|c|c|c|}
\hline 
$x$ & 0 & 1 & 2 & 3 & 4 & 5\tabularnewline
\hline
\hline 
$x_{1}^{'}$ & 0 & 0 & 0 & 1 & 1 & 1\tabularnewline
\hline 
$x_{2}^{'}$ & 0 & 1 & 2 & 0 & 1 & 2\tabularnewline
\hline
\end{tabular}
\par\end{centering}

\caption{DAT representation of $x=(x_{1}^{'},x_{2}^{'})$}

\end{table}
Consequently, the permutation matrix $A$ corresponding to Eq. \eqref{eq:A operator}
(using the standard basis for $|x\rangle$) is:

\begin{equation}
A=\left(\begin{array}{cccccc}
1 & 0 & 0 & 0 & 0 & 0\\
0 & 0 & 1 & 0 & 0 & 0\\
0 & 0 & 0 & 0 & 1 & 0\\
0 & 1 & 0 & 0 & 0 & 0\\
0 & 0 & 0 & 1 & 0 & 0\\
0 & 0 & 0 & 0 & 0 & 1\end{array}\right).\end{equation}
Also, it can be written in a compact way as $A=(0)(1,3,4,2)(5)$.
This means that $A$ leaves the coordinate states $|0\rangle$ and
$|5\rangle$ unchanged, $|1\rangle$ goes into $|3\rangle$, $|3\rangle$
goes into $|4\rangle$, $|4\rangle$ goes into $|2\rangle$ and $|2\rangle$
goes into $|1\rangle$. We note that the operator $A$ is unitary:

\begin{equation}
AA^{\dagger}=I.\label{eq:perm pr}\end{equation}

With the permutation operator $A$ at hand, we can define the new
factorization. To do this we use the pairs of operators from Eq. \eqref{eq:Zak operators},
whose definition captures a general factorization of $M=M_{1}M_{2}$.
We modify the (a) set of operators by the permutation operator $A$
of Eq. \eqref{eq:A operator}, and for convenience relabel all the
operators:\\
\begin{equation}
\begin{split}(a')\left\{ \begin{array}{l}
\tilde{U}_{1}=\tau'(\frac{2\pi{}}{a})=A\tau(\frac{2\pi{}}{a})A^{\dag}=AU^{M_{2}}A^{\dag};\\
\tilde{V}_{2}=T'(a)=AT(a)A^{\dag}=AV^{M_{1}}A^{\dag};\end{array}\right.(b)\left\{ \begin{array}{l}
\tilde{U}_{2}=\tau(\frac{2\pi{}}{b})=e^{i\hat{x}\frac{2\pi{}}{b}}=U^{M_{1}};\\
\tilde{V}_{1}=T(b)=e^{\frac{i}{\hbar}\hat{p}b}=V^{M_{2}}.\end{array}\right.\end{split}
\label{eq:New Fact}\end{equation}
The tilde denotes the new version of operators. The (a') and (b) sets
of operators replace the Schwinger operators of Eq. \eqref{eq:Sch U V}.
As we will show shortly, they obey all the commutation relations Eqs.
(\ref{eq:Schw U V 1}, \ref{eq:Schw U V 2}) of factorized operators.
Therefore, the (a') and (b) sets of operators define the new factorization,
not restricted to coprime decomposition. The commutation relation
Eq. \eqref{eq:Schw U V 2} is fulfilled by the tilde operators (Eq.
\eqref{eq:New Fact}) due to the unitarity property of $A$. For the
commutation relation Eq. \eqref{eq:Schw U V 1} we first calculate
the operation of $AU^{M_{2}}A^{\dag}$ and $A^{\dag}U^{M_{1}}A$ on
coordinate states:

\begin{equation}
\begin{split}\begin{split}\begin{array}{c}
AU^{M_{2}}A^{\dag}|x\rangle=AU^{M_{2}}A^{\dag}|x_{1}M_{2}+x_{2}\rangle=AU^{M_{2}}|x_{1}+M_{1}x_{2}\rangle=\\
=e^{\frac{2\pi i}{M_{1}}x_{1}}A|x_{1}+M_{1}x_{2}\rangle=e^{\frac{2\pi i}{M_{1}}x_{1}}|x_{1}M_{2}+x_{2}\rangle;\end{array}\end{split}
\end{split}
\end{equation}

\begin{equation}
\begin{split}\begin{array}{c}
A^{\dag}U^{M_{1}}A|x\rangle=A^{\dag}U^{M_{1}}A|x_{1}^{'}+M_{1}x_{2}^{'}\rangle=A^{\dag}U^{M_{1}}|x_{1}^{'}M_{2}+x_{2}^{'}\rangle=\\
=e^{\frac{2\pi i}{M_{2}}x_{2}^{'}}A^{\dag}|x_{1}^{'}M_{2}+x_{2}^{'}\rangle=e^{\frac{2\pi i}{M_{2}}x_{2}^{'}}|x_{1}^{'}+M_{1}x_{2}^{'}\rangle,\end{array}\end{split}
\end{equation}
where the only difference in the above calculations is that we used
different DAT based representations for the $x$ values. Using the
above expressions one can prove:

\begin{equation}
\begin{split}\begin{array}{c}
\tilde{V}_{1}^{n_{1}}\tilde{U}_{1}^{k_{1}}|x\rangle=V^{M_{2}n_{1}}AU^{M_{2}k_{1}}A^{\dag}|x_{1}M_{2}+x_{2}\rangle=e^{\frac{2\pi i}{M_{1}}k_{1}x_{1}}V^{M_{2}n_{1}}|x_{1}M_{2}+x_{2}\rangle=\\
=e^{\frac{2\pi i}{M_{1}}k_{1}x_{1}}|(x_{1}-n_{1})M_{2}+x_{2}\rangle,\end{array}\end{split}
\end{equation}
whereas applying $\tilde{U}_{1}^{k_{1}}\tilde{V}_{1}^{n_{1}}$ we
get:

\begin{equation}
\begin{split}\begin{array}{c}
\tilde{U}_{1}^{k_{1}}\tilde{V}_{1}^{n_{1}}|x\rangle=AU^{M_{2}k_{1}}A^{\dag}V^{M_{2}n_{1}}|x_{1}M_{2}+x_{2}\rangle=AU^{M_{2}k_{1}}A^{\dag}|(x_{1}-n_{1})M_{2}+x_{2}\rangle=\\
=e^{\frac{2\pi i}{M_{1}}k_{1}(x_{1}-n_{1})}|(x_{1}-n_{1})M_{2}+x_{2}\rangle.\end{array}\end{split}
\end{equation}
Similar results can be shown for the operators $\tilde{V}_{2}^{n_{2}}$
and $\tilde{U}_{2}^{k_{2}}$. Summarizing the results, the commutation
relation (Eq. \eqref{eq:Schw U V 1}) is fulfilled:

\begin{equation}
\tilde{V}_{i}^{n_{i}}\tilde{U}_{i}^{k_{i}}=\tilde{U}_{i}^{k_{i}}\tilde{V}_{i}^{n_{i}}e^{\frac{2\pi i}{M_{i}}n_{i}k_{i}},\, i=1,2.\end{equation}
Therefore, using the permutation operator $A$ we obtained the new
factorization of the unitary operators, which is not limited to coprime
decomposition of $M$. Here we used the permutation operator $A$
for the transformation of the (a) set of operators to the new (a')
set for the factorization. Obviously we could have applied the transformation
to the (b) set, which would have also enabled the factorization. 

In the particular case of $M_{1}=M_{2}$ (a=b), the two \textit{kq}
bases Eq. \eqref{eq:kq} are identical, and so are the two (a) and
(b) sets of operators in Eq. \eqref{eq:Zak operators}, and the permutation
operator $A$ from Eq. \eqref{eq:A operator} satisfies $A^{2}=I$.
In this case we have only one set of \textit{kq} - operators and only
one $|k,q\rangle$ basis. Application of the permutation operator
$A$ to that set of operators defines the tilde set, which obeys the
proper commutation relations with the original set (Eqs. \ref{eq:Schw U V 1},
\ref{eq:Schw U V 2}). Their respective eigenstates (obtained by applying
A to the unique $|k,q\rangle$ basis) are MUB with respect to the
original $|k,q\rangle$ states. (See also the next example and the
treatment of Harper - like Hamiltonians for $M=4=2^{2}$ in section
4).

The permutation operator $A$, based on the analogy to the Cooley
and Tukey FFT, solves the unitary operator factorization. To acquire
some physical intuition about the operator $A$, let us consider the
example of dimension $M=4$, where $M_{1}=M_{2}=2$. In this case,
the operators from Eq. \eqref{eq:Zak operators} in the coordinate
representation may be presented as:

{\small \begin{equation}
\begin{split}(a)\left\{ \begin{array}{l}
U^{2}=\left(\begin{array}{cccc}
1 & 0 & 0 & 0\\
0 & -1 & 0 & 0\\
0 & 0 & 1 & 0\\
0 & 0 & 0 & -1\end{array}\right),\,\, V^{2}=\left(\begin{array}{cccc}
0 & 0 & 1 & 0\\
0 & 0 & 0 & 1\\
1 & 0 & 0 & 0\\
0 & 1 & 0 & 0\end{array}\right).\end{array}\right.\end{split}
\end{equation}
}The sets (a) and (b) of operators in Eq. \eqref{eq:Zak operators}
are identical in our example. To get the second set of factorized
operators we write the (a') set from Eq. \eqref{eq:New Fact}:

{\small \begin{equation}
\begin{split}(a')\left\{ \begin{array}{l}
AU^{2}A^{\dag}=\left(\begin{array}{cccc}
1 & 0 & 0 & 0\\
0 & 1 & 0 & 0\\
0 & 0 & -1 & 0\\
0 & 0 & 0 & -1\end{array}\right),\,\, AV^{2}A^{\dag}=\left(\begin{array}{cccc}
0 & 1 & 0 & 0\\
1 & 0 & 0 & 0\\
0 & 0 & 0 & 1\\
0 & 0 & 1 & 0\end{array}\right),\end{array}\right.\end{split}
\end{equation}
}where the permutation operator in coordinate representation is $A=(0)(1,2)(3)$.
The operator $U^{2}$ has two eigenvalues (1 and -1), and the operator
$V^{2}$ permutes between the vectors with the same eigenvalue (1
or -1). This is why the operator $U^{2}$ commutes with the operator
$V^{2}$. Permutation by the operator $A$ turns the operator $U^{2}$
into the operator $AU^{2}A^{\dag}$, which anticommutes with $V^{2}$.
The operators $AU^{2}A^{\dag}$ and $V^{2}$ form a complementary
pair of operators for sub-dimension $M_{1}=2$, where their anticommutation
is consistent with Eq. \eqref{eq:Schw U V 1}. The operators $U^{2}$
and $AV^{2}A^{\dag}$ form another complementary pair of operators
for sub-dimension $M_{2}=2$. The operator $A$ permutes the eigenvalues
of $U^{2}$ in such a way as to make the operator $V^{2}$ anticommute
with $AU^{2}A^{\dag}$.

A more interesting example is dimension $M=12$, where both coprime
and non-coprime factorizations are possible. For the case of $M_{1}=2$
and $M_{2}=6$, using the coordinate representation, the permutation
operator is

\[
A=(0)(1,6,3,7,9,10,5,8,4,2)(11).\]
In the other case, where $M_{1}=3$ and $M_{2}=4$, the permutation
operator is

\[
A=(0)(1,4,5,9,3)(2,8,10,7,6)(11).\]
In both cases the construction of the operators from Eq. \eqref{eq:New Fact}
leads to the factorized pairs of operators. The generality of the
new factorization enables to perform it for every factorized numbers
$M_{1}$ and $M_{2}$. In the case where $M_{1}$ or $M_{2}$ are
composite numbers, another factorization can be performed until we
reach prime numbers in factorization.

Note that Schwinger's solution for non-coprime factorization in ref.\cite{Sch 1}
gives the factorized pairs of operators (see also ref.\cite{Sch-Eng}),
which obey the commutation relations of Eqs. (\ref{eq:Schw U V 1},
\ref{eq:Schw U V 2}). However, while for the coprime factorization
an explicit expression is given in ref.\cite{Sch 1} connecting between
the factorized pairs and the original operators $U$ and $V$, no
such expression is given for the non-coprime case. In our paper this
explicit expression is given in Eq. \eqref{eq:New Fact}.

\section{New \textit{kq}-like bases}

Each set (a') and (b) of operators (Eq. \eqref{eq:New Fact}) generates
$M$ commuting operators and can be used for the definition of a basis
for the $M$ - dimensional Hilbert space. The set (b) of operators
has, as an eigenbasis, the $|K,Q\rangle$ basis. As a result of the
unitary transformation of the (a) set, the (a') set defines the \textit{kq}-like
basis $\widetilde{|k,q\rangle}$ as follows:

\begin{equation}
\begin{split}\begin{array}{c}
(a')\,\text{ }\widetilde{|k,q\rangle}=A|k,q\rangle=\frac{1}{\sqrt{M_{2}}}\sum_{s=0}^{M_{2}-1}e^{iksa}|s+qM_{2}\rangle,\\
\\(a')\left\{ \begin{array}{l}
k=\frac{2\pi}{M}f,\,\text{ }f=0,...,M_{2}-1,\\
q=0,...,M_{1}-1.\end{array}\right.\end{array}\end{split}
\label{eq:new kq}\end{equation}
As a result of the fact that the two sets of operators $(\tilde{V}_{1},\tilde{U}_{1})$
and $(\tilde{V}_{2},\tilde{U}_{2})$ describe $M_{1}$ and $M_{2}$
subspaces in the entire $M$ - dimensional Hilbert space, the bases
$|\widetilde{k,q}\rangle$ and $|K,Q\rangle$ are mutually unbiased.
Let us check the overlap between $|\widetilde{k,q}\rangle$ and $|K,Q\rangle$
states ($a=M_{1},\, b=M_{2})$:

\begin{equation}
\langle\widetilde{k,q}|K,Q\rangle=\frac{1}{\sqrt{M_{1}}}\frac{1}{\sqrt{M_{2}}}\sum_{s=0}^{M_{2}-1}\sum_{t=0}^{M_{1}-1}e^{-iksa}e^{iKtb}\langle s+qM_{2}|Q+tb\rangle.\end{equation}
Inserting ($a=M_{1},b=M_{2})$ we have:

\begin{equation}
\begin{split}\begin{array}{c}
=\frac{1}{\sqrt{M}}\sum_{s=0}^{M_{2}-1}\sum_{t=0}^{M_{1}-1}e^{-iksM_{1}}e^{iKtM_{2}}\langle s+qM_{2}|Q+tM_{2}\rangle=\\
\\=\frac{1}{\sqrt{M}}\sum_{s=0}^{M_{2}-1}\sum_{t=0}^{M_{1}-1}e^{-iksM_{1}}e^{iKtM_{2}}\delta^{M_{1}}(s-Q)\delta^{M_{2}}(q-t)=\\
\\=\frac{1}{\sqrt{M}}e^{-ikQM_{1}}e^{iKqM_{2}}.\end{array}\end{split}
\end{equation}
Here $\delta^{M_{i}}(x-x_{0})$ means that the argument of the delta
function is taken modulo $M_{i},$ $\delta^{M_{i}}(0)=1$ and elsewhere
is zero. Therefore, these bases are mutually unbiased: $|\langle\widetilde{k,q}|K,Q\rangle|^{2}=\frac{1}{M}$.
We call the basis $|\widetilde{k,q}\rangle$ a \textit{kq}-like basis
because it is not an eigenfunction of the same operators as the $|k,q\rangle$
basis, but of the operators related to them by the permutation transformation.
In addition, it has different periodicity properties. As $|\widetilde{k,q}\rangle$
is defined, it has the completely periodic property:

\begin{equation}
(a')\,|\widetilde{k+\frac{2\pi}{M_{1}},q}\rangle=|\widetilde{k,q+M_{1}}\rangle=|\widetilde{k,q}\rangle.\end{equation}

To show explicitly the difference and similarity between the bases
$|\widetilde{k,q}\rangle$ and $|k,q\rangle$ we consider an example
of dimension $M=6$ with $M_{1}=3$ and $M_{2}=2$. Using the $|x\rangle$
representation we list in three columns all basis members of $|k,q\rangle$
on the left hand side, all $|\widetilde{k,q}\rangle$ basis vectors
in the middle and $|K,Q\rangle$ on the right hand side:

{\footnotesize \begin{equation}
\begin{split}\left\{ \begin{array}{l}
|(0,0)\rangle=\frac{1}{\sqrt{2}}\left(|0\rangle+|3\rangle\right),\\
|(0,1)\rangle=\frac{1}{\sqrt{2}}\left(|1\rangle+|4\rangle\right),\\
|(0,2)\rangle=\frac{1}{\sqrt{2}}\left(|2\rangle+|5\rangle\right),\\
|(1,0)\rangle=\frac{1}{\sqrt{2}}\left(|0\rangle-|3\rangle\right),\\
|(1,1)\rangle=\frac{1}{\sqrt{2}}\left(|1\rangle-|4\rangle\right),\\
|(1,2)\rangle=\frac{1}{\sqrt{2}}\left(|2\rangle-|5\rangle\right),\end{array}\right.\left\{ \begin{array}{c}
|\widetilde{(0,0)}\rangle=\frac{1}{\sqrt{2}}\left(|0\rangle+|1\rangle\right),\\
|\widetilde{(0,1)}\rangle=\frac{1}{\sqrt{2}}\left(|2\rangle+|3\rangle\right),\\
|\widetilde{(0,2)}\rangle=\frac{1}{\sqrt{2}}\left(|4\rangle+|5\rangle\right),\\
|\widetilde{(1,0)}\rangle=\frac{1}{\sqrt{2}}\left(|0\rangle-|1\rangle\right),\\
|\widetilde{(1,1)}\rangle=\frac{1}{\sqrt{2}}\left(|2\rangle-|3\rangle\right),\\
|\widetilde{(1,2)}\rangle=\frac{1}{\sqrt{2}}\left(|4\rangle-|5\rangle\right),\end{array}\right.\left\{ \begin{array}{c}
|(0,0)\rangle=\frac{1}{\sqrt{3}}\left(|0\rangle+|2\rangle+|4\rangle\right),\,\,\,\,\,\,\,\,\,\,\,\,\,\,\,\,\,\,\,\,\,\,\,\,\\
|(0,1)\rangle=\frac{1}{\sqrt{3}}\left(|1\rangle+|3\rangle+|5\rangle\right),\,\,\,\,\,\,\,\,\,\,\,\,\,\,\,\,\,\,\,\,\,\,\,\,\\
|(1,0)\rangle=\frac{1}{\sqrt{3}}\left(|0\rangle+e^{\frac{2\pi i}{3}}|2\rangle+e^{\frac{4\pi i}{3}}|4\rangle\right),\\
|(1,1)\rangle=\frac{1}{\sqrt{3}}\left(|1\rangle+e^{\frac{2\pi i}{3}}|3\rangle+e^{\frac{4\pi i}{3}}|5\rangle\right),\\
|(2,0)\rangle=\frac{1}{\sqrt{3}}\left(|0\rangle+e^{\frac{4\pi i}{3}}|2\rangle+e^{\frac{2\pi i}{3}}|4\rangle\right),\\
|(2,1)\rangle=\frac{1}{\sqrt{3}}\left(|1\rangle+e^{\frac{4\pi i}{3}}|3\rangle+e^{\frac{2\pi i}{3}}|5\rangle\right).\end{array}\right.\end{split}
\end{equation}
}Hence, the $|\widetilde{k,q}\rangle$ and the $|k,q\rangle$ bases
are neither equal nor orthogonal to one another (they are eigenfunctions
of different sets of operators). Nevertheless, both these bases are
mutually unbiased to the $|K,Q\rangle$ basis in this coprime case.

\section{Engineering of the energy spectrum using the permutation operator
A in Harper-like Hamiltonians}

The new factorized pairs of operators $\left(T'(a),\tau(\frac{2\pi}{b})\right)$
and $\left(T(b),\tau'(\frac{2\pi}{a})\right)$ (Eq. \eqref{eq:New Fact})
describe $M_{2}$ and $M_{1}$ dimensional subspaces, respectively
\cite{Sch 1}. Therefore, replacing the Harper-like Hamiltonian $H[T(b),\tau(\frac{2\pi}{a})]$
of ref.\cite{MRZ} by $H[T'(b),\tau(\frac{2\pi}{a})]$ we should obtain
$M_{1}$ energy levels, each of which is degenerate $M_{2}$ times,
without restriction for $M_{2}$ and $M_{1}$ to be coprime.

To show the advantage of the new factorization, we compare it with
the energy spectra design method of ref. \cite{MRZ}. As a first example,
let us consider the dimension $M=6$. We choose the simple Harper-like
Hamiltonian proposed in ref. \cite{MRZ}:

\begin{equation}
\begin{split}\begin{array}{l}
H=H(T(b),\tau(\frac{2\pi}{a}))=V_{1}cos(\frac{b}{\hbar}\hat{p})+V_{2}cos(\frac{2\pi}{a}\hat{x}),\end{array}\end{split}
\label{eq:Harper H}\end{equation}
where $V_{1}$ and $V_{2}$ are constants. We solve this Hamiltonian
using the kq-representation: 

\begin{equation}
|\psi\rangle=\sum_{k,q}|k,q\rangle\langle k,q|\psi\rangle=\sum_{k,q}C_{k,q}|k,q\rangle.\end{equation}
The resulting eigenvalue equation for our Hamiltonian is:

\begin{equation}
\begin{split}\begin{array}{l}
[V_{1}cos(\frac{b}{\hbar}\hat{p})+V_{2}cos(\frac{2\pi}{a}\hat{x})]\sum_{k,q}C_{k,q}|k,q\rangle=\varepsilon\sum_{k,q}C_{k,q}|k,q\rangle.\end{array}\end{split}
\end{equation}
After applying the operators we have:

\begin{equation}
\begin{split}\begin{array}{l}
\sum_{k,q}C_{k,q}[\frac{V_{1}}{2}(|k,q-b\rangle+|k,q+b\rangle)+V_{2}cos(\frac{2\pi}{a}q)|k,q\rangle]=\varepsilon\sum_{k,q}C_{k,q}|k,q\rangle.\end{array}\end{split}
\label{eq:Harper 6}\end{equation}
The above eigenvalue equation can be solved for each value of $k=\frac{2\pi}{M}f$
independently. So for our particular choice of dimension $M=6$ with
$M_{1}=a=2$ and $M_{2}=b=3$, performing the summation over $q$
values with the use of the quasi-periodicity property of the $|k,q\rangle$
states, we obtain the following equation (for some particular $k$
value): \begin{equation}
\begin{split}\begin{array}{l}
C_{k,0}[\frac{V_{1}}{2}(e^{4ki}|k,1\rangle+e^{-2ki}|k,1\rangle)+V_{2}cos(\frac{2\pi}{2}\cdot0)|k,0\rangle]+\\
\\C_{k,1}[\frac{V_{1}}{2}(e^{2ki}|k,0\rangle+e^{-4ki}|k,0\rangle)+V_{2}cos(\frac{2\pi}{2}\cdot1)|k,1\rangle]=\\
\\=\varepsilon[C_{k,0}|k,0\rangle+C_{k,1}|k,1\rangle].\end{array}\end{split}
\end{equation}
Using the orthogonality of the $|k,q\rangle$ states the above equation
is equivalent to the solution of the following $M_{1}=2$ coupled
equations: \begin{equation}
\begin{split}\begin{array}{l}
\left(\begin{array}{cc}
V_{2} & V_{1}e^{-\frac{2\pi i}{6}4f}\\
V_{1}e^{\frac{2\pi i}{6}4f} & -V_{2}\end{array}\right)\left(\begin{array}{c}
C_{k,0}\\
C_{k,1}\end{array}\right)=\varepsilon\left(\begin{array}{c}
C_{k,0}\\
C_{k,1}\end{array}\right).\end{array}\end{split}
\end{equation}
The energy spectrum $\varepsilon$ is: \begin{equation}
\begin{split}\begin{array}{l}
\varepsilon_{1,2}=\pm\sqrt{V_{1}^{2}+V_{2}^{2}},\end{array}\end{split}
\end{equation}
which is $f$ independent and therefore each energy level is 3-fold
degenerate (note that for current example $f=\{0,1,2\}$ and $k=\frac{2\pi}{6}f$).
The relation between the coefficients is:

\begin{equation}
C_{k,0}=\frac{V_{1}e^{-\frac{2\pi i}{6}4f}}{\varepsilon-V_{2}}C_{k,1}\,\text{ or equally }\, C_{k,1}=\frac{V_{1}e^{\frac{2\pi i}{6}4f}}{\varepsilon+V_{2}}C_{k,0}.\label{eq:C k-dep phase}\end{equation}

On the other hand, if instead of coprime factorized $M=6$ we choose
$M=4$ and substitute $M_{1}=a=2$ and $M_{2}=b=2$ into equation
\eqref{eq:Harper 6}, we get Eq. \eqref{eq:Non degen} with non-degenerate
energy spectrum:

\begin{equation}
\sum_{k,q}C_{k,q}[\frac{V_{1}}{2}(|k,q-2\rangle+|k,q+2\rangle)+V_{2}cos(\frac{2\pi}{2}q)|k,q\rangle]=\varepsilon\sum_{k,q}C_{k,q}|k,q\rangle.\label{eq:Non degen}\end{equation}
Using the quasi-periodicity properties of $|k,q\rangle$ states we
have:

\begin{equation}
\sum_{k,q}C_{k,q}[V_{1}cos(2k)|k,q\rangle+V_{2}cos(\frac{2\pi}{2}q)|k,q\rangle]=\varepsilon\sum_{k,q}C_{k,q}|k,q\rangle,\end{equation}
and the energy spectrum is:

\begin{equation}
\varepsilon_{1,2,3,4}=\pm V_{1}\pm V_{2}.\label{eq:E non-dege}\end{equation}
This result is expected, because of the absence of factorization into
sub-dimensions $M_{1}=2$ and $M_{2}=2$ using the operators of Eq.
\eqref{eq:Zak operators}. 

Let us now follow the same procedure with the new operators of Eq.
\eqref{eq:New Fact}. Accordingly, the Harper-like Hamiltonian of
Eq. \eqref{eq:Harper H} changes to: \begin{equation}
\begin{split}\begin{array}{l}
H=H(T(b),\tau'(\frac{2\pi}{a}))=V_{1}cos(\frac{b}{\hbar}\hat{p})+V_{2}Acos(\frac{2\pi}{a}\hat{x})A^{\dag}.\end{array}\end{split}
\label{eq:new Harper H}\end{equation}
To compare the two schemes we solve the above Hamiltonian using the
$\widetilde{kq}$-representation: 

\begin{equation}
|\psi\rangle=\sum_{k,q}|\widetilde{k,q}\rangle\langle\widetilde{k,q}|\psi\rangle=\sum_{k,q}\widetilde{C}_{k,q}|\widetilde{k,q}\rangle.\end{equation}
The eigenvalue equation for our Hamiltonian is: \begin{equation}
\begin{split}\begin{array}{l}
[V_{1}cos(\frac{b}{\hbar}\hat{p})+V_{2}Acos(\frac{2\pi}{a}\hat{x})A^{\dag}]\sum_{k,q}\widetilde{C}_{k,q}|\widetilde{k,q}\rangle=\varepsilon\sum_{k,q}\widetilde{C}_{k,q}|\widetilde{k,q}\rangle.\end{array}\end{split}
\end{equation}
After applying the operators we have:

\begin{equation}
\begin{split}\begin{array}{l}
\sum_{k,q}\widetilde{C}_{k,q}[\frac{V_{1}}{2}(|\widetilde{k,q-1}\rangle+|\widetilde{k,q+1}\rangle)+V_{2}cos(\frac{2\pi}{a}q)|\widetilde{k,q}\rangle]=\varepsilon\sum_{k,q}\widetilde{C}_{k,q}|\widetilde{k,q}\rangle,\end{array}\end{split}
\label{eq:new Harper 6}\end{equation}
where we have used the two relations:

\[
T(b)|\widetilde{k,q}\rangle=|\widetilde{k,q-1}\rangle\text{ and }\tau'(\frac{2\pi}{a})|\widetilde{k,q}\rangle=e^{\frac{2\pi i}{a}q}|\widetilde{k,q}\rangle.\]
As before, the eigenvalue equation \eqref{eq:new Harper 6} can be
solved for each value of $k$ independently, and using the complete
periodicity property of the $|\widetilde{k,q}\rangle$ we have (with
$M_{1}=a=2$ and $M_{2}=b=3$): \begin{equation}
\begin{split}\begin{array}{l}
\tilde{C}_{k,0}[\frac{V_{1}}{2}(|\widetilde{k,1}\rangle+|\widetilde{k,1}\rangle)+V_{2}cos(\frac{2\pi}{2}\cdot0)|\widetilde{k,0}\rangle]+\\
\\\tilde{C}_{k,1}[\frac{V_{1}}{2}(|\widetilde{k,0}\rangle+|\widetilde{k,0}\rangle)+V_{2}cos(\frac{2\pi}{2}\cdot1)|\widetilde{k,1}\rangle]=\\
\\=\varepsilon[\tilde{C}_{k,0}|\widetilde{k,0}\rangle+\tilde{C}_{k,1}|\widetilde{k,1}\rangle].\end{array}\end{split}
\end{equation}
 In matrix form the above equation reads: \begin{equation}
\begin{split}\begin{array}{l}
\left(\begin{array}{cc}
V_{2} & V_{1}\\
V_{1} & -V_{2}\end{array}\right)\left(\begin{array}{c}
\tilde{C}_{k,0}\\
\tilde{C}_{k,1}\end{array}\right)=\varepsilon\left(\begin{array}{c}
\tilde{C}_{k,0}\\
\tilde{C}_{k,1}\end{array}\right).\end{array}\end{split}
\end{equation}
 Hence, we get the same spectrum of energies as before, with each
level being 3-fold degenerate: \begin{equation}
\begin{split}\begin{array}{l}
\varepsilon_{1,2}=\pm\sqrt{V_{1}^{2}+V_{2}^{2}},\end{array}\end{split}
\end{equation}
 and a new relation between the coefficients:

\begin{equation}
\tilde{C}_{k,0}=\frac{V_{1}}{\varepsilon-V_{2}}\tilde{C}_{k,1}\,\text{ or equally }\,\tilde{C}_{k,1}=\frac{V_{1}}{\varepsilon+V_{2}}\tilde{C}_{k,0}.\end{equation}
In comparison with the previous solution Eq. \eqref{eq:C k-dep phase},
now we do not have a k-dependent phase in the relation between the
coefficients $\tilde{C}_{k,0}$ and $\tilde{C}_{k,1}$.

In the case of $M=4$, solving equation \eqref{eq:new Harper 6} with
$M_{1}=a=2$ and $M_{2}=b=2$, we have:

\begin{equation}
\sum_{k,q}\widetilde{C}_{k,q}[\frac{V_{1}}{2}(|\widetilde{k,q-1}\rangle+|\widetilde{k,q+1}\rangle)+V_{2}cos(\frac{2\pi}{2}q)|\widetilde{k,q}\rangle]=\varepsilon\sum_{k,q}\widetilde{C}_{k,q}|\widetilde{k,q}\rangle.\end{equation}
As a result of the $k$ independence of the equation above it is equivalent
to the eigenvalue equation considered for dimension $M=6$. Therefore
(as one can easily check) we have to solve the matrix equation:

\begin{equation}
\begin{split}\begin{array}{l}
\left(\begin{array}{cc}
V_{2} & V_{1}\\
V_{1} & -V_{2}\end{array}\right)\left(\begin{array}{c}
\tilde{C}_{k,0}\\
\tilde{C}_{k,1}\end{array}\right)=\varepsilon\left(\begin{array}{c}
\tilde{C}_{k,0}\\
\tilde{C}_{k,1}\end{array}\right),\end{array}\end{split}
\end{equation}
and consequently the corresponding energy levels, with each level
being 2-fold degenerate, are: \begin{equation}
\begin{split}\begin{array}{c}
\varepsilon_{1,2}=\pm\sqrt{V_{1}^{2}+V_{2}^{2}}.\end{array}\end{split}
\end{equation}
Therefore, in spite of the non-degenerate spectrum of the Hamiltonian
$H(T(b),\tau(\frac{2\pi}{a}))$ for non-coprime $M_{1}$ and $M_{2}$,
for the Hamiltonian $H(T(b),\tau'(\frac{2\pi}{a}))$ the energy levels
preserve their degeneracies.

\section{Summary and discussion}

The main result of our work is a generalization of the Schwinger unitary
operator factorization to non-coprime factorizations. That is, for
a composite dimension $M=M_{1}M_{2}$, we factorize the $U$ and $V$
operators from Eq. \eqref{eq:UV} into two pairs of operators $(\tilde{U}_{1},\tilde{V}_{1})$
and $(\tilde{U}_{2},\tilde{V}_{2})$ Eq. \eqref{eq:New Fact}. Each
of the pairs generates a complete orthogonal operator basis for the
sub-dimensions $M_{1}$ and $M_{2}$, and operators from different
bases commute. The factorization enables us to consider any single
physical system with dimension $M=M_{1}M_{2}$ as a pair of physical
systems in $M_{1}$ and $M_{2}$ - factorized degrees of freedom,
where $M_{1}$ and $M_{2}$ are not restricted to be coprime. Considering
factorized operators may simplify various $M$ - dimensional phase
space problems in the same way as the Cooley-Tukey FFT simplifies
the application of the DFT. Moreover, the new factorization deepens
our physical intuition. In particular, we applied the new factorization
to a Harper-like Hamiltonian model, and developed an algorithm for
energy spectrum design in this model. Using the algorithm, we can
construct a Hamiltonian, which is a function of the operators $(\tilde{U}_{1},\tilde{V}_{1})$.
Therefore, it is designed to obtain $M_{1}$ energy levels (with a
spectrum determined by Hamiltonian's details), each level being $M_{2}$-fold
degenerate. The algorithm of the energy spectrum design can be of
interest, for example in solid state physics for electrons in a strong
magnetic field \cite{Zak Mag}.

The application of the permutation operator $A$ (which is the key
to the solution for the non-coprime cases) to the \textit{kq} bases
problem generates the \textit{kq}-like basis which is a MUB to the
original $|K,Q\rangle$ basis. This \textit{kq}-like basis has a different
periodicity property than the original \textit{kq} bases: it is completely
periodic in the coordinate and momentum variables simultaneously.

\section*{Acknowledgments}

AM is grateful to Prof. Wei-Min Zhang for his very kind hospitality
in Tainan. The work of AM was partly supported by grant HUA97-12-02-161
at NCKU. The authors acknowledge numerous informative and helpful
discussions with Professor Michael Revzen.

\end{document}